\documentclass[aps,prl,twocolumn,superscriptaddress]{revtex4-1}
\usepackage{amssymb, amsmath,  hyperref, graphicx, bm, color}
\begin{document}
\newcommand {\mb} {\mu_B}
\newcommand {\tpc} {T_\mathrm{pc}}
\newcommand {\tc} {T_c^0}
\newcommand {\ms} {m_s^\mathrm{phys}}
\newcommand {\ml} {m_l^\mathrm{phys}}
\newcommand {\g} {f_G}
\newcommand {\F} {F_\mathrm{reg}}
\newcommand {\zc} {z_c}
\newcommand {\todo}[1] {\textcolor{red}{#1}}


\title{Universality driven analytic structure of the QCD crossover: radius of convergence
in the baryon chemical potential}




\author{Swagato\ Mukherjee}
\affiliation{Physics Department, Brookhaven National Laboratory, Upton, NY 11973,
USA}

\author{Vladimir\ Skokov}
\affiliation{Department of Physics, North Carolina State University, Raleigh, NC 27695, USA}
\affiliation{RIKEN/BNL Research Center, Brookhaven National Laboratory, Upton, NY 11973}

\begin{abstract}

Recent lattice QCD calculations strongly indicate  that the chiral crossover of
QCD at zero baryon chemical potential (\(\mb\)) is a remnant of the  second order
chiral phase transition. Universal properties of this second order phase transition can be mapped to 
QCD temperature \(T\) and \(\mb\) using non-universal parameters determined by lattice QCD recently. Motivated by
these results, first, we discuss the analytic structure of the partition
function in the QCD crossover regime --- the so-called Yang-Lee edge singularity --- 
solely based on universal properties. Next, utilizing the lattice QCD results for
non-universal parameters we map this singularity to the real \(T\) and complex
\(\mb\) plane, leading to the determination of  the radius of convergence in \(\mb\)
in the QCD crossover regime. These universality- and QCD-based results
provide tight constraints on the range of validity of the lattice QCD calculations at
\(\mb>0\).  Implication of this result on the location of the conjectured QCD critical
point  is discussed.

\end{abstract}
\date{\today}
\maketitle
\section{\label{sc:1} Introduction}
The chiral symmetry of quantum chromodynamics (QCD) is spontaneously broken in the
vacuum. First-principle lattice QCD calculations have conclusively shown that the
approximate chiral symmetry  with physical values of quark masses gets nearly
restored at a pseudo-critical temperature \(\tpc=156.5 \pm
1.5\)~MeV~\cite{Bazavov:2018mes} via a smooth crossover~\cite{Bhattacharya:2014ara,
Bazavov:2011nk}. Lattice QCD calculations have also shown that similar chiral
symmetry restoring crossover takes place  at
small-to-moderate values of \(\mb\); the dependence of \(\tpc(\mb)\) has been
computed~\cite{Bazavov:2018mes}. It has been {\it conjectured} that at some sufficiently
large values of \(\mb\) the chiral restoration in QCD takes place through a first
order transition; the point in the \(T-\mb\) phase diagram at which the chiral
crossover line turns into a first order phase transition line is known as the QCD
critical end point (CEP), for a review, see Ref.~\cite{Fukushima:2010bq}.

While experimental searches to locate this conjectured QCD critical point are
on-going at RHIC and SPS, presently, first-principle lattice QCD calculations only
provide very limited guidance on the existence and location of the QCD critical point
in the \(T-\mb\) phase diagram, because a direct lattice QCD calculation at
\(\mb\ne0\) is hindered by the fermion sign problem. The present lattice calculations
providing information on the QCD thermodynamics --- either by carrying out Taylor
expansions around \(\mb=0\)~\cite{Bazavov:2017dus} or through analytic continuation
from purely imaginary values of \(\mb\)~\cite{Bonati:2015bha, Bellwied:2015rza,
Borsanyi:2018grb} --- crucially rely on the assumption that the QCD partition
function is an analytic function of complex \(\mb\) within a radius of convergence.
To what extent these lattice QCD results are trustworthy, and how far in \(\mb\)
these methods might be extended can be answered only if we have reliable knowledge of
the radius of convergence of the QCD partition function  around \(\mb=0\).  

{
In this work, we extract this radius of convergence   
for the first time based on universal properties of QCD phase transition and first principal 
non-universal input from lattice QCD. 
}
To that end, we will provide an estimate for the location of the
singularity nearest to \(\mb=0\) and for \(T \sim \tpc\), {based on the
universal behavior of the QCD partition function with {
nearly} massless u/d-quarks and {
{\it mapping} 
this universal structure to QCD \(T\) and \(\mu\) plane using non-universal} input from (lattice) QCD calculations. 

The universal analytic structure of
the QCD crossover and its connection to the radius of convergence in
complex-\(\mb\) plane were previously discussed in Refs.~\cite{Stephanov:2006dn,
Almasi:2019bvl}, and a mean-field (random matrix) model-based estimate for the radius
of convergence also was provided in  Ref.~\cite{Stephanov:2006dn}.}

{
	The main idea is as follows. A system near a second order phase transition typically falls into one of the limited number of the 
	universality classes.
	The universality class is fully defined by the global symmetries (O(4) symmetry for the chiral phase transition in QCD) and the number of spatial dimensions ($d$ = 3 for QCD). The corresponding universal equation of state is a function of two variables: 
	i) The so-called reduced temperature, $t$, that does not explicitly break the symmetry of the system. The parameter $t$ measures the deviation from the second order phase transition point in the phase diagram. For finite-temperature QCD, $t$ is a combination of $T$ and $\mu_B$ (both parameters do not explicitly break O(4) symmetry).  
	ii) The so-called magnetic field $h$ -- the relevant parameter that breaks the symmetry explicitly. In QCD, for the chiral phase transition, $h$ simply corresponds to the light quark mass. 
A true second order (chiral) phase transition takes place at $t=h=0$, while  for any fixed $h\ne0$ the system undergoes a smooth
crossover transition as a function of the parameter $t$. This is manifest in the dependence of the equation of state on $(t, h)$ in the real domain 
(see e.g. Ref.~\cite{Engels:2011km}).  
The apparent smoothness of the crossover equation of state disguises the singularity of the partition function at complex values of $(t, h)$ --  the Yang-Lee edge singularity~\cite{Fisher:1978pf}. In the vicinity of a second order phase transition, in the scaling regime, and for a fixed value of $h$, this singularity 
limits the radius of convergence of a series expansions of thermodynamical observables in powers of $t$. Thus,
the knowledge of 
the (universal) location of the Yang-Lee edge singularity  and the (non-universal) mapping between $(t, h)$ and QCD variables $(T, \mu_B, m_l)$ 
would be sufficient in order to determine the radius of convergence. 
Recent theoretical advances in lattice QCD and Ref.~\cite{Connelly:2020gwa} provide both required ingredients and allow us to extract the radius of convergence of the Taylor series of the pressure near zero baryon chemical potential. 
}

\begin{figure}[!t]
    \centering
    \includegraphics[width=0.45\textwidth]{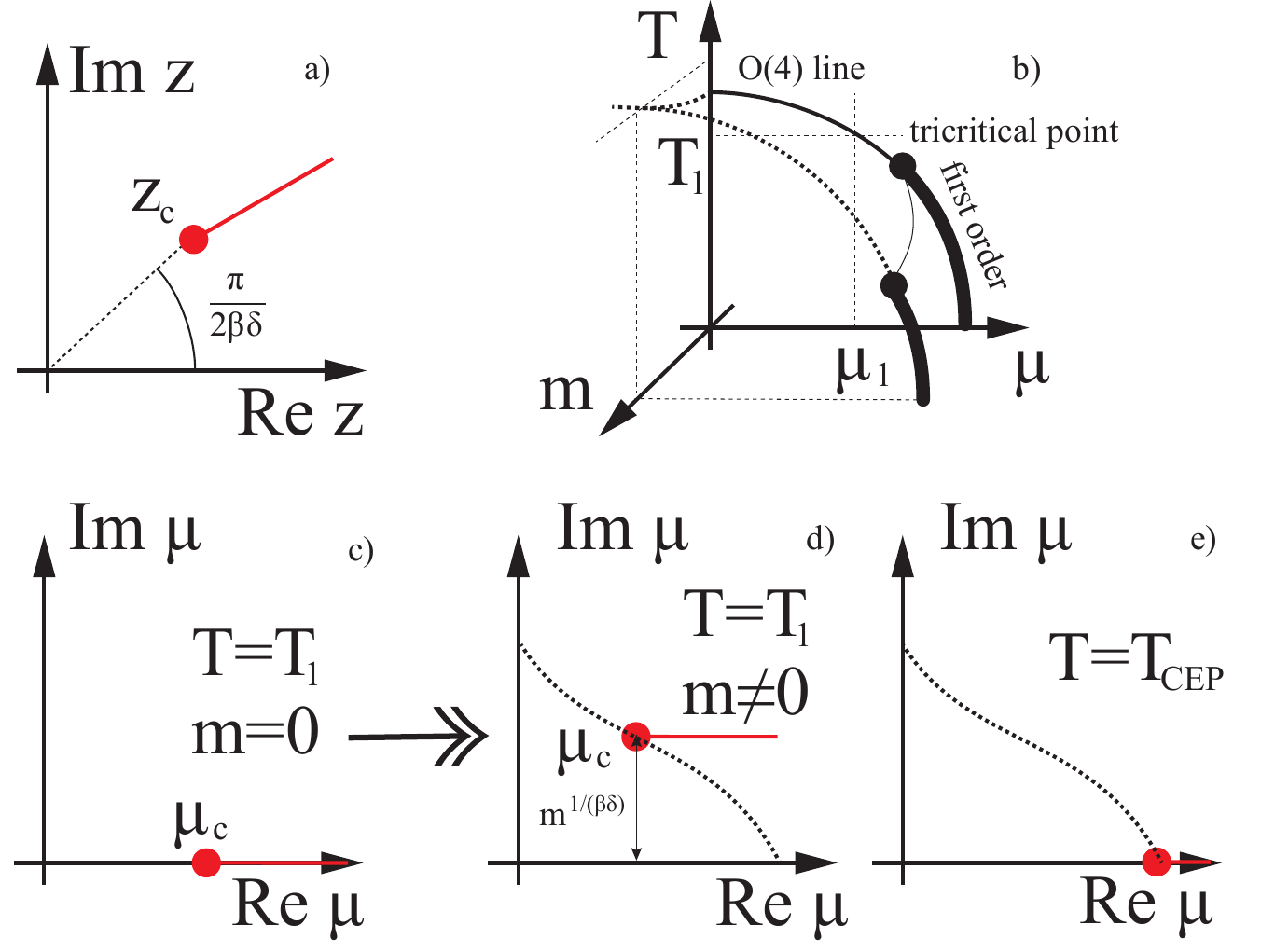}
    \caption{Illustration of analytic structure of the universal equation of state and its mapping to QCD:  a) Yang-Lee edge singularity of the magnetic equation of state; the argument of the singularity is defined by the O(4) critical exponents; b)  illustration of phase diagram in the plane of  temperature -- baryon chemical potential -- light quark mass; note that all three variables can be combined into a single variable $z$ in the scaling regime, see Eq.~\eqref{eq:z};  
    for the purpose of illustration, we consider an arbitrary temperature $T_1$ above the triciritical temperature in the chiral limit and not smaller than the temperature of the phase transition at zero chemical potential; 
    c) analytic structure in the complex chemical potential plane in the chiral limit for a fixed temperature $T_1>T_{\rm tricritical}$; d)  analytic structure for non-zero light quark mass for $T_1>T_{\rm CEP}$: the singularity from the real axis is shifted to the complex values of the chemical potential; this situation corresponds to the crossover phase transition;   e)  the same as d) but for  $T=T_{\rm CEP}$: the singularity and its complex conjugate (not shown) approach and pinch real axis resulting in  a critical end point (CEP).    O(4) magnetic equation of state with mapping Eq.~\eqref{eq:z} fairly describes analytical structure in the O(4) scaling regime.}  
    \label{fig:Ill}
\end{figure}

\section{\label{sc:2} Analytic structure of the QCD free energy for small quark
mass}
{
Lattice QCD provides a compelling evidence that 
for massless u/d-quarks the chiral symmetry restoration in QCD 
takes place 	 
via a `true' second order chiral phase transition} at \(\mb=0\) (for a review, see
Ref.~\cite{Fukushima:2010bq})~\footnote{One cannot completely rule out that 
	QCD demonstrates a weak fluctuation-induced first-order phase transition in the chiral limit for two massless quarks. Current lattice QCD calculations with staggered quarks strongly suggest that the transition is of the second order.  We caution the reader that the staggered fermion lattice calculations are known to break down in the chiral limit at finite lattice spacings. The conclusion on the second 
	order chiral phase transition is also consistent with Refs.~\cite{Burger:2011zc,Umeda:2016qdo,Cuteri:2018wci,
Endrodi:2018xto}. 
}. Recent progresses in lattice QCD calculations have
shown that for \(\mb=0\) the chiral phase transition of 2+1-flavor QCD, with
massless up and down quarks and a physical strange quark, takes place at the chiral
transition temperature \(\tc=132^{+3}_{-6}\)~MeV~\cite{Ding:2019prx}. The
universality class of the chiral phase transition of  2+1-flavor QCD was found to be
consistent with that of the three-dimensional \(O(4)\) spin model~\cite{Ding:2019prx,
Ding:2018auz, Ejiri:2009ac, Kaczmarek:2011zz}.

Based on the universal argument, if the light u/d-quark mass, \(m_l\), is sufficiently small,  the three dimensional QCD diagram of \(m_l\), \(\mb\), and \(T\) can be mapped onto a {\it one-dimensional} space using the so-called scaling variable, $z$.  This mapping is suitable above the possible tricritical point. The behavior of the order parameter
(the
u/d-chiral condensate up to an overall constant)
can then be described by~\cite{Kaczmarek:2011zz}
\begin{equation}
  M(T, m_l, \mb) = \left( \frac{m_l}{\ms} \right)^{\frac{1}{\delta}} \g(z)
  + \F(T, m_l, \mb) \,,
  \label{eq:M}
\end{equation}
where the scaling variable, $z$, -- a combination of three independent variables (light quark masses, $T$ and $\mb$) --    is given by
\begin{align}
  z &= z_0 \left( \frac{m_l}{\ms}\right)^{-\frac{1}{\beta\delta}} \notag \\ &\times
  \left[ \frac{T-\tc}{\tc} + \kappa^B_2 \left( \frac{\mb}{\tc} \right)^2 +  \kappa^B_4 \left( \frac{\mb}{\tc}  \right)^4  + \dots \right] \,.
  \label{eq:z}
\end{align}
The form of the scaling variable follows from the one of 
the \(O(N)\) spin model  $z=z_0 h^{-\frac{1}{\beta \delta}} t$. Here $h$ is the symmetry breaking field which in QCD corresponds to the light quark mass; while $t$ is the field responsible for the deviation from the criticality without explicit symmetry breaking; in QCD, a combination of $T-\tc$ and $\mb$ play the role of $t$. The chemical potential due to charge conjugation symmetry can contribute through even powers only.   

{
Thus the analytic properties of $\g(z)$ encode the entirety of the critical behavior in the scaling regime of  \(O(4)\)  phase transition.  That is the whole three dimensional QCD diagram in Fig.~\ref{fig:Ill}b) for $T$ about the tricritical temperature with the analytical structure in the complex chemical potential plane  in 
Fig.~\ref{fig:Ill}c) and Fig.~\ref{fig:Ill}d) can be read off from  Fig.~\ref{fig:Ill}a) and the properties of a single-variable function $\g(z)$ . 
}

In equation~\eqref{eq:M}, \(\g\) is the universal scaling function~\footnote{
This is the so-called magnetic equation of state for three-dimensional O(4)
(Z(2))
universality class if QCD has a second (first) order phase transition in the massless limit for u/d-quarks. The mapping~\eqref{eq:z} was established for the second order phase transition in the chiral limit. Our analysis however can be extended to a first order phase transition case, if lattice QCD provides non-universal input on mapping parameters.  },  \(\beta\) and \(\delta\) are the universal critical
exponents;  while \(z_0\) and the curvature parameters of \(\tpc(\mb)\),
\(\kappa^B_{2,4}\), are non-universal. Current lattice
QCD calculations established   that \(\kappa^B_4\) is consistent with zero  within the
precision of the calculation~\cite{Bazavov:2018mes}; based on this we set
\(\kappa^B_4\) and possible higher order corrections denoted by ellipses to zero in
what follows. The physical strange quark mass is denoted by \(\ms\). {For
simplicity, we only consider the leading singular contribution in Eq.~\eqref{eq:M}. 
There are also sub-leading scaling corrections (see e.g. Ref.~\cite{amit2005field}).
However,  they 
do not modify the location of the
singularities in \(\{T, \mb, m\}\)-plane and, therefore, are not important in the context of our work.}
The function \(\F\) characterizes small deviations, if any, from the scaling
behavior. Lattice calculations have provided evidence that chiral observables of QCD
with physical values of u/d-quark masses are well-described by Eq.~\eqref{eq:M} by
including only small corrections from \(\F\)~\cite{Ding:2019prx, Li:2017aki,
Ejiri:2009ac, Kaczmarek:2011zz}. Since \(\F\) is not expected to have any
singularities close to zero chemical potential~\footnote{We can definitively state that 
the so-called thermal singularities originating from zeroes of the inverse Fermi-Dirac function will not modify our analysis as they lead to the radius of convergence $R/T<\pi$.}, the analyticity of \(M\) in the
complex-\(\mb\) plane is governed by the analytic structure of the universal function
\(\g\). The corresponding universal chiral behavior of the QCD free energy is given
by \((m_l/\ms)^{1+1/\delta} f(z)\), where the scaling function \(f(z)\) is related to
\(\g\) through: \( \g = zf^\prime(z)/(\beta\delta) - (1+ 1/\delta) f (z) \); the
prime denotes the derivative with respect to \(z\).

It is well-known that in the complex-\(z\) plane the function \(\g\) has a
singularity; it is  the so-called Yang-Lee edge singularity~\footnote{It comes with its complex conjugate pair; this is implicitly assumed in the remaining text}
of the form \(\g \sim
(z-\zc)^\sigma\)~\cite{Fisher:1978pf}, see Fig.~\ref{fig:Ill}a).
 The corresponding most singular contribution to the free energy is  then 
\(f(z) \sim (z-\zc)^{1+\sigma}\). {
 Since there has been some confusion in the QCD literature, we comment} that  the critical exponent for this singularity of the free energy,
\(1+\sigma\) is not related to the $O(N)$ specific heat critical exponent,
\(2-\alpha\), as one naively might assume. Additionally, the critical exponent
\(\phi\) introduced in Ref.~\cite{Almasi:2019bvl} for the crossover singularities is
nothing but \(\sigma\). The Yang-Lee edge singularity can be treated as an ordinary
critical point, belonging to the $Z(2)$ universality class of  $\phi^3$ theory in three
spatial dimension, and with a purely imaginary coupling~\cite{Fisher:1978pf}. {
 Although, as long as it is positive,  the exact value of $\sigma$ is not important for our analysis, we want to point out that it has been known for more than four decades -- for all
finite values of \(N\) of three-dimensional \(O(N)\) universality class  \(\sigma=0.085(1)\)~\cite{Gliozzi:2014jsa, Fisher:1978pf}.   The
argument of \(\zc\) is known in terms of the \(O(N)\) critical
exponents~\cite{Itzykson:1983gb} as a consequence of the Lee-Yang theorem   \(\zc = |\zc|  e^{ i
\frac{\pi}{2\beta\delta}}\) as illustrated in Fig.~\ref{fig:Ill}a). 
In contrast to the exponent and the argument of $z_c$, the absolute value $|z_c|$ was not know until recently, see Ref.~\cite{Connelly:2020gwa}.  
  We stress that  owing to the universality of \(\g\) as a function of
(complex) \(z\) the value of \(|\zc|\) is also universal. Analytically, \(\zc\) can be
calculated in two limits--- mean-field and \(N \to \infty\).}

\section{\label{sc:3} Analytic structure of \(\g\) in the mean-field and \(N \to \infty\) limits}

In both limits, the mean-field and the \(N\to\infty\), \(\g\) can be represented  in  the following general
form~\cite{PhysRevB.7.1967}
\begin{equation}
  \g \left[ z + \g^2  \right]^\gamma =1 \,.
  \label{eq:g}
\end{equation}
The specific cases can be obtained by plugging in the corresponding critical
exponents --- \(\gamma=1\) for mean-field, and \(\gamma=2\) for \(N\to\infty\).
For the mean-field case this equation can also be
obtained  straightforwardly by varying the  free energy of the \(\phi^4\) Landau-Ginzburg
theory~\cite{Almasi:2016gcn} with respect to the order parameter \(\phi\). To determine the Yang-Lee edge singularity
branch-point, \(\zc\), from \(\g\) we consider the inverse function \(z(\g)\). The
branch-point can be obtained from the condition \((dz/d\g)_{\zc}=0\). 
 Considering only the branch closest to \(z=0\) on the physical Riemann sheet, i.e., the one connected to $z=0$ as defined by the
standard normalization condition \(\g(0)=1\),  Eqs.~\eqref{eq:g} and  \((dz/d\g)_{\zc}=0\) 
completely determine
\begin{align}
  	\zc^{\rm MF}  = \frac{3}{2^{2/3}} e^{ i\frac{\pi}{3}}
     \quad \quad \text{mean-field} \,, \nonumber \\
    \zc^{N\to\infty}  = \frac{5}{2^{8/5}} e^{ i\frac{\pi}{5}}
    \quad  \quad N \to \infty \,.
    \label{eq:zc}
\end{align}

\section{\label{sc:4} Radius of convergence in the complex-\(\mb\) plane}

\begin{figure}[!t]
    \centering
    \includegraphics[width=0.45\textwidth]{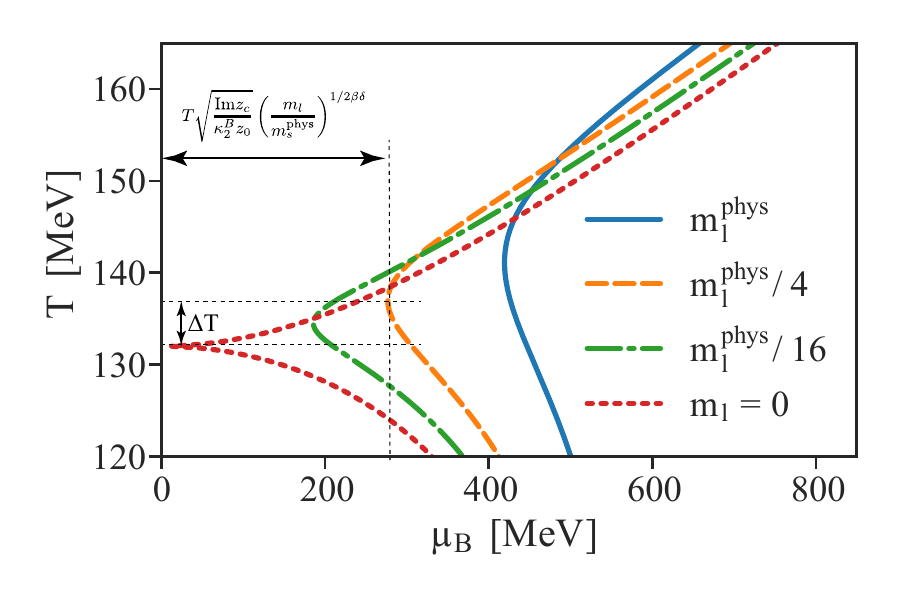}
    \caption{Radius of convergence \(\mb\) for different values of $T$ and for different values of the
    light up/down quark masses. The minimum of the curves shifts to higher temperatures  
    by the amount $\frac{\Delta T}{T^0_c} = \frac{{\rm Re} z_c} {z_0}  \left( \frac{ m_l }{ \ms} \right)^{\frac{1}{\beta\delta}}$. See text for details.   
}
    \label{fig:Rm}
\end{figure}

As evident from  Eqs.~\eqref{eq:M} and \eqref{eq:z},  for a  fixed value of \(m_l>0\),
the derivatives of \(M\) with respect to \(\mb\) are proportional to the
derivatives \(\g(z)\) with respect to \(z\). Thus, the convergence of the Taylor
expansion in \(\mb\) around \(\mb=0\), as well as analytic continuation in the
complex-\(\mb\) plane are bounded by the value of \(\zc\). Specifically, from Eq.~\eqref{eq:z} for a definite value of
$T$ the Taylor series about zero chemical potential will have the radius of
convergence given by  
\begin{equation}
	R_{\rm conv} = \left|\frac{\zc}{z_0}
\left(\frac{m_l}{\ms}\right)^{1/\beta\delta}  - \frac{T-\tc}{\tc} \right|^{1/2}
\frac{T}{\sqrt{\kappa^B_2}}\,.
\label{Eq:Rconv} 
\end{equation}
 At $T=\tc$, the radius of convergence is directly proportional to
\(|\zc|\). 
In general, $R_{\rm conv}$   depends not only on the magnitude of \(\zc\) but also on its phase. 
From this equation it follows, that the radius of convergence assumes its minimal value
$T \sqrt{\frac{ {\rm Im} \zc}{ \kappa^B_2 z_0}}
\left(\frac{m_l}{\ms}\right)^{1/2\beta\delta}$
when the real part of the first term cancels completely the second term in
Eq.~\eqref{Eq:Rconv}.  This minimum shifts to temperatures  higher then $T_c^0$
    by the amount $\frac{\Delta T}{T^0_c} = \frac{{\rm Re} z_c} {z_0}  \left( \frac{ m_l }{ \ms} \right)^{\frac{1}{\beta\delta}}$.   
	As expected both values are defined by \(\zc\), see also Fig.~\ref{fig:Rm}. 
	
The previous discussion clearly demonstrate that, in the scaling regime, 
validity of the Taylor expansions in \(\mb\) and the analytic
continuations in complex-\(\mb\) plane of the QCD free energy is determined by  the
value of \(|\zc|\).  2+1-flavor lattice QCD calculations show
that the chiral condensate, \(M\), for the physical value of light up/down quark
mass, \(\ml=\ms/27\), are well-described by the 3-dimensional \(O(4)\) scaling
function \(\g\), with  inclusion of small corrections from the analytic function
\(\F\)~~\cite{Ding:2019prx, Li:2017aki, Ejiri:2009ac, Kaczmarek:2011zz}. Obviously,
\(\F\) unavoidably affects  the values of the low-order Taylor coefficients; however, 
any analytic contribution does not change the radius of convergence. Base on these arguments, we expect that, for QCD, the singularity nearest to \(\mb=0\) in
the complex-\(\mb\) plane is defined by \(\zc\). If \(\zc\) is known then
Eq.~\eqref{eq:z} can be used to translate this singularity to the complex-\(\mb\)
plane and, thereby, determine the corresponding radius of convergence.  The rest of
the universal and non-universal parameters entering Eq.~\eqref{eq:z} are known--- (i)
The critical exponents of the \(O(4)\) universality class \(\beta=0.380\),
\(\delta=4.824\)~\cite{Engels:2003nq}. (ii) Both \(\ml\) and \(T\) are purely real.
(iii) \(\tc=132^{+3}_{-6}\)~MeV~\cite{Ding:2019prx}. (iv) The curvature of the
pseudo-critical temperature \(\tpc(\mb)\),
\(\kappa_2^B=0.012(2)\)~\cite{Bazavov:2018mes}. (v) {The scale factor \(z_0\)
can be determined by fitting the \(m_l\)-dependence of the lattice QCD-calculated
\(\tpc(m_l)\)~\cite{Bazavov:2011nk};} based on the lattice QCD results of
Ref.~\cite{Ding:2019prx} on \(\tpc(m_l)\)  the scale factor is estimated to be \(z_0
\simeq 1-2\)~\footnote{We thank Anirban Lahiri for providing us this input on behalf
of the HotQCD collaboration. Recent results of the HotQCD collaboration on the value of \(z_0\) can also be found in Ref.~\cite{Clarke:2020htu}.}.
Currently the best estimate for \(|\zc|\)  is available from
the Functional Renormalization Group studies~\cite{Connelly:2020gwa}; they show that
\(|\zc|\approx 1.665\)  for \(O(4)\). This value is accidentally  close to the one obtained in the large $N$ limit  $\approx 1.649$. 
In our analysis we use \(|\zc|=1.665\). 
We note that the Functional Renormalization Group approach is well suited for extracting the location of the edge singularity, as 
it does not rely on the Monte-Carlo importance sampling and thus does not suffer from the sign problem which hinders lattice simulation at complex (imaginary) values of $z$ ($h$).  Moreover, critical behavior is dominated  by the long-range physics of the slow critical modes. This justifies the  derivative expansion of the effective action;  in the non-perturbative Functional Renormalization Group approach, this expansion is know to rapidly converge \cite{Balog:2019rrg}.  
To account for a possible systematic uncertainty of the truncation scheme used in the Functional Renormalization Group
calculation we generously  vary \(|\zc|\) by 5\%. The large-$N$ value of  \(|\zc|\) 
falls into this uncertainty band. 

In Figure~\ref{fig:Rm} we show the radius of convergence in \(\mb\) in the \(T-\mb\)
plane for different values of \(m_l\) in the range \(0-\ml\), using  \(z_0=2\), \(O(4)\) critical exponents, and other lattice QCD-determined
non-universal parameters described above. Note that, in the chiral limit, QCD free energy
is singular at \(T=\tc\), \(\mb=0\) and, therefore, the radius of
 convergence at this point is zero, see also Refs.~\cite{Stephanov:2006dn, Almasi:2019bvl}.

\begin{figure}[!t]
    \centering
    \includegraphics[width=0.45\textwidth]{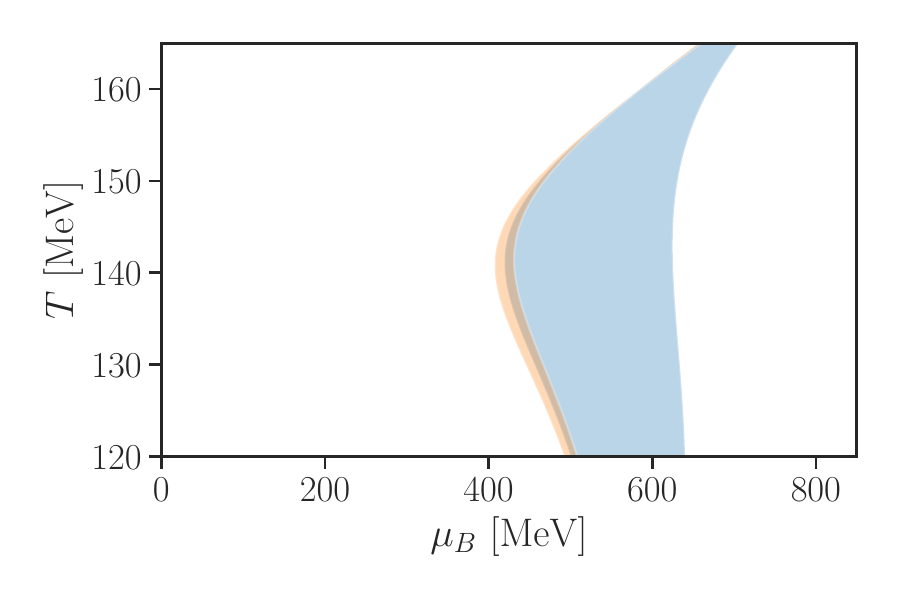}
    \caption{Radius of convergence in \(\mb\) for physical quark masses. The orange
    band is for {\(z_0=2\)} and incorporates a 5\% uncertainty on the value of \(|\zc|\). The blue
    band depicts variation of \(z_0=1-2\).  }
    \label{fig:R}
\end{figure}

Figure~\ref{fig:R} provides a more realistic estimate for the radius of convergence
in \(\mb\) in the \(T-\mb\) plane for \(\ml\) by varying \(|\zc|\) around its
FRG value and \(z_0=1-2\). While the value of
\(|\zc|\) was recently determined to rather high precision and leads to a limited uncertainty of the radius of convergence, more precise
lattice QCD result for \(z_0\) is needed to improve this estimate.

\section{\label{sec:5} Conclusions}

Relying only on the universal behavior of QCD in the chiral crossover region we
investigated the analytic behavior of the free energy. We argued that if the chiral
behavior of QCD is well-described by the universal scaling, as borne out in recent
the lattice QCD calculations, then the analytic structure  of the free energy will be completely
governed by the corresponding universal scaling function. We estimated the relevant singularity of the scaling
function based on the two extreme limits of mean-field and \(N\to\infty\). For the analysis, we used 
the location of the singularity extracted in Functional Renormalization Group calculations. We showed
how this can be translated to the singularity in the complex-\(\mb\) plane
to determine the radius of convergence in \(\mb\). Our results are solely based on the universal
input  and well-determined non-universal parameters from lattice QCD
calculations. Figure~\ref{fig:R} summarizes our universality- and QCD-based
estimate for the radius of convergence in \(\mb\) for temperatures in the vicinity of
the QCD chiral crossover. It shows that the radius of convergence is larger than
\(|\mb| \gtrsim 400\)~MeV, implying that the present lattice QCD calculations based on
Taylor expansions in \(\mb\) and analytic continuations from imaginary values of
\(\mb\) can be reliable below this region, as suggested also by recent lattice QCD
calculations~\cite{Bazavov:2018mes, Bazavov:2017dus, Borsanyi:2019hsj}.

The current state-of-the-art lattice QCD calculations do not find any evidence for an
additional singularity for \(\mb \lesssim 400\)~MeV~\cite{Bazavov:2018mes,
Bazavov:2017dus,  Borsanyi:2019hsj}. Our result on the radius of convergence
\(|\mb|\gtrsim400\)~MeV,  coupled with these lattice QCD results, suggest that QCD
critical point, if one exists, will most likely be located at \(\mb\gtrsim400\)~MeV.


\section*{Acknowledgments}

  This material is based upon work supported by the U.S. Department of Energy, Office
  of Science, Office of Nuclear Physics: (i) Through the Contract No. DE-SC0012704;
  (ii) Through the contract No. DE-SC0020081;
  (iii) Within the framework of the Beam Energy Scan Theory (BEST) Topical
  Collaboration.

  V.S. also thanks the ExtreMe Matter Institute EMMI (GSI Helmholtzzentrum f\"ur
  Schwerionenforschung, Darmstadt, Germany) for partial support and their
  hospitality.

  We thank Bengt Friman, Frithjof Karsch, Lex Kemper, {Misha Stephanov,} Robert
  Pisarski, Krzysztof Redlich,  Thomas Sch\"afer, and Mithat \"Unsal for illuminating
  discussions.

  We thank the organizers of EMMI Workshop ``Probing the Phase Structure of Strongly
  Interacting Matter: Theory and Experiment'',  which inspired us to work together on
  this project.

\bibliography{ref}
\end{document}